\documentclass[12pt,english]{article}
\usepackage[T1]{fontenc}
\usepackage[latin1]{inputenc}
\usepackage{amsmath}
\usepackage{graphicx}
\usepackage{babel}

\providecommand{\LyX}{L)\kern-.1667em\lower.25em\hbox{Y}\kern-.125emX\@}
\makeatother

\begin{document}

\title{\textbf{Electrodynamics in Friedmann-Robertson-Walker Universe: Maxwell and
Dirac fields in Newman-Penrose formalism }}
\author{\textbf{U. Khanal}
\footnote{khanalu@yahoo.com}
\\Central Department of Physics, Tribhuvan University, Kirtipur, \textbf{Nepal}}
\date{\today}
\maketitle

\begin{abstract}
Maxwell and Dirac fields in Friedmann-Robertson-Walker spacetime is
investigated using the Newman-Penrose method. The variables are all
separable, with the angular dependence given by the spin-weighted spherical
harmonics. All the radial parts reduce to the barrier penetration problem,
with mostly repulsive potentials representing the centrifugal energies. Both
the helicity states of the photon field see the same potential, but that of
the Dirac field see different ones; one component even sees attractive
potential in the open universe. The massless fields have the usual
exponential time dependencies; that of the massive Dirac field is coupled to
the evolution of the cosmic scale factor $a$. The case of the radiation
filled flat universe is solved in terms of the Whittaker function. A formal
series solution, valid in any FRW universe, is also presented. The energy
density of the Maxwell field is explicitly shown to scale as $a^{-4}$. The
co-moving particle number density of the massless Dirac field is found to be
conserved, but that of the massive one is not. Particles flow out of certain
regions, and into others, creating regions that are depleted of certain
linear and angular momenta states, and others with excess. Such current of
charged particles would constitute an electric current that could generate a
cosmic magnetic field. In contrast, the energy density of these massive
particles still scales as $a^{-4}$.
\end{abstract}

\pagebreak

\section{Introduction}

The Newman-Penrose\cite{NP} formalism of projecting vectors, tensors and
spinors onto a set of null tetrad bases has proved to be an immensely useful
tool with which to investigate the properties of quantum fields in curved
spacetime. It has been used successfully in various black hole geometries 
\cite{chandra1,lohia,uk}. The method has also been used more\ recently\cite
{mukhopadhyay} to further study the behaviour of Dirac particles in
different geometries. As the geometry of our Universe is of the
Friedmann-Robertson-Walker type, it is important to study how
electrodynamics is altered by the real expanding Universe in contrast to
that in Minkowskian spacetime. The behaviour of the quantum fields in FRW
spacetime must be indicative of the exact nature of the geometry of our
Universe, particularly whether it is flat, closed or open. Properties of
matter fields in general, and the massive Dirac fields in particular, must
have profound consequences on the structure formation process. Propagation
of electromagnetic waves in FRW spacetime was studied by Haghighipour\cite
{nader} (see references therein), among other authors.The Dirac field was
studied in the NP formalism by Zecca\cite{zecca} and the work was continued
further by Sharif\cite{sharif}. This paper is another attempt to use the NP
method to study the behaviour of Maxwell and Dirac fields in the Universe. A
slightly different null-tetrad bases are chosen in this paper, resulting in
equations that take on the familiar barrier penetration forms without much
ado. The Maxwell equations for free electromagnetic waves are exactly
solvable for all cases in terms of well known special functions. The spatial
part of the Dirac equations are also exactly solvable; the time evolution
however is coupled to the evolution of the scale factor, can be solved in
terms of known functions only for the case of a radiation dominated universe.

The FRW line element is usually written in the comoving coordinates as\cite
{weinberg} 
\begin{equation}
ds^{2}=dt^{2}-a^{2}(t)\left( \frac{dr^{\prime 2}}{1-Kr^{\prime 2}}+r^{\prime
2}\left( d\theta ^{2}+\sin ^{2}\theta d\phi ^{2}\right) \right) ,
\label{1.1}
\end{equation}
where $a(t)$ is the scale factor; the curvature constant $K$ can be scaled
to the values $0,+1$ or $-1$ to describe the flat, closed or open universe
respectively. In terms of the conformal time $\eta =\int \frac{dt}{a}$, and
a change of variable $r^{\prime }=r/\left( 1+Kr^{2}/4\right) =r/f(r)$, the
line element Eq. (\ref{1.1}) becomes\cite{adler} 
\begin{equation}
ds^{2}=a^{2}(\eta )\left[ d\eta ^{2}-\frac{1}{f^{2}(r)}\left(
dr^{2}+r^{2}\left( d\theta ^{2}+\sin ^{2}\theta d\varphi ^{2}\right) \right) %
\right] .  \label{1.2}
\end{equation}
Following Chandrasekhar's guidelines for the black hole\cite{chandra}, we
can introduce the null tetrad $l_{\mu }=\left( 1,-\frac{1}{f},0,0\right)
,n_{\mu }=\frac{a^{2}}{2}\left( 1,\frac{1}{f},0,0\right) ,m_{\mu }=\frac{ar}{%
\sqrt{2}f}\left( 0,0,-1,-i\sin \theta \right) $, and the complex conjugate $%
m_{\mu }^{\ast }$ . The non vanishing spin-coefficients are $\beta =-\alpha =%
\frac{f}{2\sqrt{2}ar}\cot \theta ,\rho =-\frac{f}{a^{2}}\left( \frac{\dot{a}%
}{af}-\frac{\left( f/r\right) ^{\prime }}{f/r}\right) ,\mu =\frac{f}{2}%
\left( \frac{\dot{a}}{af}+\frac{\left( f/r\right) ^{\prime }}{f/r}\right) $,
and $\gamma =-\frac{\dot{a}}{2a},$where the prime and the overdot denote
derivatives with respect to $r$ and $\eta $ respectively.

\begin{figure}[tbp]
\includegraphics[scale=1]{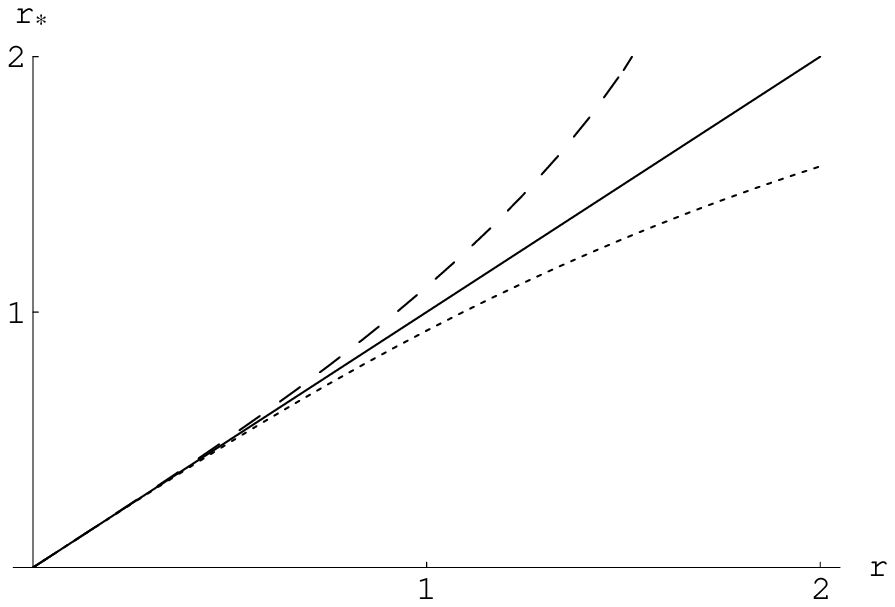}\newline
\caption{The tortoise coordinate $r_{\ast }$ as a function of $r$. In flat
space with $K=0,$ (solid), the two co-ordinates are identical. For $K=1$
(dotted), $r_{\ast }$ plateaus off to a value of $\protect\pi $ at large $r$%
. For $K=-1$ (dashed), the maximum value that $r$ can have is $2$, where $%
r_{\ast }\rightarrow \infty $. For small $r$, all three cases are alike.}
\label{tortoise}
\end{figure}

Besides the vanishing of the spin-coefficient $\kappa $ because the $l$
vector forms a congruence of null geodesics, and of $\pi $ telling us that
the tetrad bases remain unchanged as they are parallely propagated along $l$
as in Ref.\cite{zecca}, the present choice of the bases are affinely
parameterized so that $\varepsilon $ also vanishes. The directional
derivatives can then be written down as $D=l^{\mu }\partial _{\mu }=\frac{1}{%
a^{2}}D_{-},\Delta =n^{\mu }\partial _{\mu }=-\frac{1}{2}D_{+},$ $\delta
=m^{\mu }\partial _{\mu }=\frac{f}{\sqrt{2}ar}L_{-},$ and $\delta ^{\ast
}=m^{\ast \mu }\partial _{\mu }=\frac{f}{\sqrt{2}ar}L_{+},$where $D_{\pm }=f%
\frac{\partial }{\partial r}\mp \frac{\partial }{\partial \eta }$and $L_{\pm
}=\frac{\partial }{\partial \theta }\mp \frac{i}{\sin \theta }\frac{\partial 
}{\partial \varphi }$. In terms of what may be called the ``tortoise''
coordinate defined by 
\begin{equation}
r_{\ast }=\int \frac{dr}{f}=\frac{2}{\sqrt{K}}\text{arctan}\frac{\sqrt{K}r}{2%
}=\left\{ 
\begin{array}{cc}
r, & K=0 \\ 
2\,\text{arctan}(r/2), & K=1 \\ 
2\,\text{arctanh}(r/2), & K=-1
\end{array}
\right. ,  \label{1.3}
\end{equation}
$D_{\pm }=\frac{\partial }{\partial r_{\ast }}\mp \frac{\partial }{\partial
\eta }$. The tortoise coordinate shown in Fig. \ref{tortoise} reveals that,
near the origin all the $r_{\ast }$'s coincide with each other; also, from
Eq. (\ref{1.3}) it is seen that as $r\rightarrow 0,r_{\ast }\rightarrow r$
for both $K=\pm 1$; thus, at small distances both the closed and open
universes look basically flat.

The Maxwellian wave equations are solved in the next Section, and the
solutions are used to investigate the behaviour of the energy-momentum in
the expanding Universe. Section 3 is devoted to Dirac equations; the spatial
parts of the four components are easily solved, and so is the time
dependence of the massless field; the time dependence of the massive field
is more complicated as it is coupled to the evolution of $a$; nonetheless,
the case of the radiation filled flat universe is found to have a solution
in terms of the Whittaker function. A formal series solution that can be
used in any FRW universe is also written down. The behaviour of the Dirac
current and energy-momentum is then investigated. The final Section makes
some concluding remarks.

\section{Maxwell Field}

\begin{figure}[tbp]
\includegraphics[scale=1]{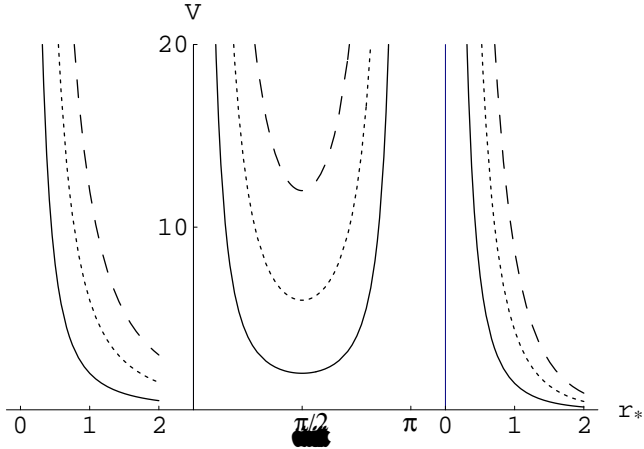} \newline
\caption{The potential seen by electromagnetic waves. $l=1,2$, and $3$, for
the solid, dotted and dashed lines. $K=0,1$, and $-1$ from left to right.
The potential is a $1/r^{2}$ wall in the first, become infinitely high wall
at each boundary in the second, and an exponentially decaying wall in the
third case. The walls are thicker for higher $l$.}
\label{empot}
\end{figure}

In the NP formalism, the six independent components of the antisymmetric
Maxwell tensor $F_{\mu \nu }$ are replaced by the three complex scalars\cite
{chandra} $\Phi _{0}=F_{\mu \nu }l^{\mu }m^{\nu },\ \Phi _{1}=\frac{1}{2}%
F_{\mu \nu }\left( l^{\mu }n^{\nu }+m^{\ast \mu }m^{\nu }\right) ,\,\Phi
_{2}=F_{\mu \nu }m^{\ast \mu }n^{\nu }$. Then the Maxwell equations become 
\begin{eqnarray}
\left( D-2\rho \right) \Phi _{1} &=&\left( \delta ^{\ast }+\pi -2\alpha
\right) \Phi _{0}-\kappa \Phi _{2},  \notag \\
\left( \delta -2\tau \right) \Phi _{1} &=&\left( \Delta +\mu -2\gamma
\right) \Phi _{0}-\sigma \Phi _{2},  \notag \\
\left( D-\rho +2\varepsilon \right) \Phi _{2} &=&\left( \delta ^{\ast }+2\pi
\right) \Phi _{1}-\lambda \Phi _{0}\,\text{and}  \notag \\
\left( \delta -\tau +2\beta \right) \Phi _{2} &=&\left( \Delta +2\mu \right)
\Phi _{1}-\nu \Phi _{0}.  \label{2.1}
\end{eqnarray}
With the spin-coefficients and the directional derivatives from Sec. 1, and
using the fact that $(1/f)D_{\pm }\left[ a^{n}(f/r)^{m}\Phi \right]
=a^{n}(f/r)^{m}\left( \frac{1}{f}D_{\pm }\mp \,\frac{n\dot{a}}{af}+m\frac{%
\left( f/r\right) ^{\prime }}{\left( f/r\right) }\right) ,$ Eqs. (\ref{2.1})
become 
\begin{eqnarray}
D_{-}\left[ \sqrt{2}\Phi _{1}\left( \frac{ar}{f}\right) ^{2}\right]
&=&\left( L_{+}+\cot \theta \right) \left[ \Phi _{0}\frac{ra^{3}}{f}\right] ,
\notag \\
L_{-}\left[ \sqrt{2}\Phi _{1}\left( \frac{ar}{f}\right) ^{2}\right] &=&-%
\frac{r^{2}}{f^{2}}D_{+}\left[ \Phi _{0}\frac{ra^{3}}{f}\right] ,  \notag \\
\frac{r^{2}}{f^{2}}D_{-}\left[ \frac{2ar}{f}\Phi _{2}\right] &=&L_{+}\left[ 
\sqrt{2}\Phi _{1}\left( \frac{ra}{f}\right) ^{2}\right] \,\text{and}  \notag
\\
\left( L_{-}+\cot \theta \right) \left[ \frac{2ar}{f}\Phi _{2}\right]
&=&-D_{+}\left[ \sqrt{2}\Phi _{1}\left( \frac{ra}{f}\right) ^{2}\right] .
\label{2.2}
\end{eqnarray}
The first of the two pairs of Eqs. (\ref{2.2}) can be used together to
eliminate $\Phi _{1}$ and be combined into a second order equation for $\Phi
_{0}$ only, and similarly from the second pair for the equation for $\Phi
_{2}$ only. Writing $\Phi _{+}=\Phi _{0}ra^{3}/f$ and $\Phi _{-}=2\Phi
_{2}ar/f,$the resulting two equations take the form 
\begin{equation}
D_{\mp }\left[ \frac{r^{2}}{f^{2}}D_{\pm }\Phi _{\pm }\right] +L_{\mp }\left[
L_{\pm }+\cot \theta \right] \Phi _{\pm }=0.  \label{2.3}
\end{equation}
It is clear that the Maxwell scalars in Eq. (\ref{2.3}) are separable into $%
\Phi _{\pm }=\phi _{\pm }(r,\eta )Y_{\pm 1}(\theta ,\varphi )$. Then the
angular parts become 
\begin{equation}
\left( \frac{1}{\sin \theta }\frac{\partial }{\partial \theta }\sin \theta 
\frac{\partial }{\partial \theta }\pm \frac{2i\cos \theta }{\sin ^{2}\theta }%
\frac{\partial }{\partial \varphi }+\frac{1}{\sin ^{2}\theta }\frac{\partial
^{2}}{\partial \varphi ^{2}}-\cot ^{2}\theta -1\right) Y_{\pm 1}=-\lambda
Y_{\pm 1},  \label{2.4}
\end{equation}
where $\lambda $ is the separation constant. Same eigenvalue $\lambda $ is
used for both the solutions $Y_{\pm 1}$because the equation for $Y_{+1}$
goes over to that for $Y_{-1}$ with the change of variable $\theta
\rightarrow \pi -\theta ;$ the two just represent the two helicity states of
the photon. The solutions are the spin weighted spherical harmonics of
Goldberg, et al. \cite{goldberg} with $\lambda =l(l+1);$ $l$ is the total
angular momentum of the field including its spin, so in this case for
photons, $l=1,2,3,...$. It is obvious that the azimuthal part is just $%
e^{im\varphi }$. The radial and time parts then become 
\begin{equation}
D_{\mp }\left[ \frac{r^{2}}{f^{2}}D_{\pm }\phi _{\pm }\right] =\lambda \phi
_{\pm }.  \label{2.5}
\end{equation}
The relative normalization between the two helicity states is set by the
easily verifiable Teukolsky type relations 
\begin{equation}
D_{\pm }\left[ \frac{r^{2}}{f^{2}}D_{\pm }\phi _{\pm }\right] =\lambda \phi
_{\mp }.  \label{2.6}
\end{equation}
Substituting $Z=$ $\frac{r^{2}}{f^{2}}D_{\pm }\phi _{\pm }$ for the
dependent variable in Eq. (\ref{2.5}), operating on it with $D_{\pm }$, and
writing $D^{2}=D_{\pm }D_{\mp }=\frac{\partial ^{2}}{\partial r_{\ast }^{2}}-%
\frac{\partial ^{2}}{\partial \eta ^{2}},$ we end up with the equation of
motion$D^{2}Z=VZ,$ with the potential $V=\lambda (f/r)^{2}$ independent of $%
\eta $. Consequently, the time dependence $e^{-i\omega \eta }$ can be
factored out. For the stipulated time and azimuthal dependence, $D_{\pm }=%
\frac{\partial }{\partial r_{\ast }}\pm i\omega $ and $L_{\pm }=\frac{%
\partial }{\partial \theta }\pm \frac{m}{\sin \theta }$. Then the radial
parts of $\phi _{\pm }$, which may be written as $R_{\pm }$ in Eq. (\ref{2.5}%
), satisfy complex conjugate equations. Finally, we are left with the
equation for the radial dependence of $Z$ as, 
\begin{equation}
\left[ \frac{d^{2}}{dr_{\ast }^{2}}+\omega ^{2}\right] R=VR,  \label{2.7}
\end{equation}
which has taken the familiar form of the one dimensional quantum mechanical
barrier penetration problem with the potentials for the three cases given by 
\begin{equation}
V=\left\{ 
\begin{array}{cc}
l(l+1)/r^{2}, & K=0 \\ 
l(l+1)/\sin ^{2}r_{\ast }, & K=1 \\ 
l(l+1)/\sinh ^{2}r_{\ast }, & K=-1
\end{array}
\right. .  \label{2.8a}
\end{equation}
Eq. (\ref{2.7}) is basically the same as that derived in Ref.\cite{nader}.
The potentials barriers represented by Eq. (\ref{2.8a}) are just the
rotational kinetic energies. They become thicker for larger $l$. The
barriers go down as $1/r^{2}$ in the flat universe, fall off even more
rapidly in the open universe, and are potential wells between $0<$ $r_{\ast
}<\pi $ that rise to infinity towards both the ends in the closed universe
(Fig. \ref{empot}). If $r_{\ast }$ is replaced by $ir_{\ast }$ and $\omega $
by $i\omega $ in the potential for $K=1$ in Eq. (\ref{2.8a}) we end up with
the one for $K=-1$. Both the spin states of the photon satisfy the same
equation with the same real potential. The inversion relation is provided by
the normalization condition Eq. (\ref{2.6}) as 
\begin{equation}
\lambda R_{\pm }=D_{\mp }R.  \label{2.8}
\end{equation}

\subsection{Solutions}

All the three cases of Eq. (\ref{2.7}) can fortunately be solved. For $K=1$, 
$f(r)=1\,$and $r_{\ast }=r$. Then Eq. (\ref{2.7}) has the general solution 
\cite{stegun} 
\begin{equation}
R=r\left[ A_{0}j_{l}\left( \omega r\right) +B_{0}y_{l}\left( \omega r\right) %
\right] ,  \label{2.2.1}
\end{equation}
where $j_{l}$ and $y_{l}$ are the respective spherical Bessel functions. As $%
l\geq 1$, only the $j$'s are regular at the origin. In the asymptotic
region, as $2j_{l}=h_{l}^{(1)}+h_{l}^{(2)}\rightarrow \left[
(-i)^{l+1}e^{i\omega r}+(i)^{l+1}e^{-i\omega r}\right] /\omega $, we can use
the inversion relation Eq. (\ref{2.8}) to see that the incoming waves are
dominated by $R_{+}$and the outgoing by $R_{-}$.

In the closed universe, it is most convenient to write the solutions in
terms of the Gegenbauer functions $C_{n}^{\alpha }$ as 
\begin{equation}
R=A_{1}\left( \sin r_{\ast }\right) ^{l+1}C_{\omega -l-1}^{l+1}\left( \cos
r_{\ast }\right) +B_{1}\left( \sin r_{\ast }\right) ^{-l}C_{\omega
+l}^{-l}\left( \cos r_{\ast }\right) .  \label{2.2.2}
\end{equation}
Again, the first one is regular within the allowed range of $r_{\ast }$, and 
$\omega \geq l+1$. These em waves form standing waves.

We may write the solution in the open universe as 
\begin{eqnarray}
R &=&A_{-1}e^{i\omega r_{\ast }}\,_{2}F_{1}\left( -l,l+1,1-i\omega ,\frac{1}{%
2}(1-\coth r_{\ast })\right)  \notag \\
&&+B_{-1}e^{-i\omega r_{\ast }}\,_{2}F_{1}\left( -l,l+1,1+i\omega ,\frac{1}{2%
}(1-\coth r_{\ast })\right) ,  \label{2.2.3}
\end{eqnarray}
consisting of the outgoing and incoming parts at $r_{\ast }\rightarrow
\infty $. Alternatively, we may make the replacements $r_{\ast }\rightarrow
ir_{\ast }$ and $\omega \rightarrow i\omega $ in Eq. (\ref{2.2.2}) to study
the behaviour near the origin. Once $\Phi _{0}$ and $\Phi _{2}$ are known,
Eq. (\ref{2.1}) can be used to determine $\Phi _{1}$ and complete the
solutions for the electromagnetic field.

\subsection{Energy Momentum}

The energy-momentum tensor of the Maxwell field in terms of the $\Phi $'s is
given by 
\begin{eqnarray}
4\pi T_{\alpha \beta } &=&\left| \Phi _{0}\right| ^{2}n_{\alpha }n_{\beta
}+\left| \Phi _{2}\right| ^{2}l_{\alpha }l_{\beta }+2\left| \Phi _{1}\right|
^{2}\left\{ l_{(\alpha }n_{\beta )}+m_{(\alpha }m_{\beta )}^{\ast }\right\} 
\notag \\
&&-4\Phi _{0}^{\ast }\Phi _{1}n_{(\alpha }m_{\beta )}-4\Phi _{1}^{\ast }\Phi
_{2}l_{(\alpha }m_{\beta )}+2\Phi _{0}^{\ast }\Phi _{2}m_{(\alpha }m_{\beta
)}+\text{cc,}  \label{2.3.1}
\end{eqnarray}
where the brackets around the indices of the null-tetrad denote
symmetrization and cc is the complex conjugates of preceeding terms; $%
T_{\alpha \beta }$ is obviously traceless. We are particularly interested in
the time components which can be determined as 
\begin{eqnarray}
4\pi \rho &=&4\pi T_{\eta }^{\eta }=\frac{f^{2}}{2r^{2}a^{4}}\left[ \left|
\Phi _{+}\right| ^{2}+\left| \Phi _{-}\right| ^{2}\right] +2\left| \Phi
_{1}\right| ^{2}  \notag \\
&=&-4\pi \left( T_{r}^{r}+T_{\theta }^{\theta }+T_{\varphi }^{\varphi
}\right)  \notag \\
4\pi T_{\eta }^{r} &=&-\frac{f^{3}}{2r^{2}a^{4}}\left[ \left| \Phi
_{+}\right| ^{2}-\left| \Phi _{-}\right| ^{2}\right] ,  \label{2.3.2}
\end{eqnarray}
where $\rho $ is the energy density. The $\theta $ and $\varphi $ components
are not important for our analysis. Noting that the necessary non-vanishing
Christoffel symbols are $\Gamma _{\eta \eta }^{\eta }=\Gamma _{r\eta
}^{r}=\Gamma _{\theta \eta }^{\theta }=\Gamma _{\varphi \eta }^{\varphi }=%
\dot{a}/a$ where the overdot represents derivative with respect to $\eta $,
we may write the conservation equation as $\partial _{\eta }\left( \sqrt{g}%
T_{\eta }^{\eta }\right) =-\partial _{i}\left( \sqrt{g}T_{\eta }^{i}\right) $%
, where $g=-\det g_{\alpha \beta }=\left( \frac{a^{4}r^{2}}{f^{3}}\sin
\theta \right) ^{2}=a^{8}g_{3}$ where $g_{3}$ is the determinant of the
3-space; the terms with the $\Gamma $'s are not written as their sum vanish
due to the tracelessness. Integrating over the three space we find 
\begin{equation}
\partial _{\eta }\left( Ea^{4}\right) =-a^{4}\int \partial _{i}\left( \sqrt{%
g_{3}}T_{\eta }^{i}\right) =-a^{4}\int d\Sigma _{r}T_{\eta }^{r}=a^{4}\int
d\Sigma _{r}\frac{f^{3}}{r^{2}}\left( \left| \Phi _{+}\right| ^{2}-\left|
\Phi _{-}\right| ^{2}\right) ,  \label{2.3.3}
\end{equation}
where $E$ is the total energy contained within a spherical volume of radius $%
r$, and $d\Sigma _{r}=\frac{r^{2}}{f^{3}}\sin \theta d\theta d\varphi $ is
an elemental area normal to the radial direction. Using the inversion
relation, Eq. (\ref{2.8}), we get $\lambda ^{2}\left[ \left| \Phi
_{+}\right| ^{2}-\left| \Phi _{-}\right| ^{2}\right] =\left[ \left|
D_{-}Z\right| ^{2}\left| Y_{+}\right| ^{2}-\left| D_{+}Z\right| ^{2}\left|
Y_{-}\right| ^{2}\right] =\left[ \left| \frac{dR}{dr_{\ast }}\right|
^{2}+\omega ^{2}\left| R\right| ^{2}\right] \left( \left| Y_{+}\right|
^{2}-\left| Y_{-}\right| ^{2}\right) $\linebreak $+i\omega W[R^{\ast
},R]\left( \left| Y_{+}\right| ^{2}+\left| Y_{-}\right| ^{2}\right) $, where 
$W$ is the Wronskian. As the $Y$' s are normalized to unity, when we perform
the angular integration, the first term on the left hand side will vanish.
Now $R$ and $R^{\ast }$ satisfy the same equation and are proportional to
the same function so that the Wronskian will also vanish.Hence $Ea^{4}$
remains constant, and so does $\rho a^{4}$, during the evolution of the
Universe. As a consequence, there is no net flux of electromagnetic energy
flowing out of a closed surface. This is a direct proof of the well known
result that the energy density of massless radiation scales as $a^{-4}$.

\section{Dirac Field}

In NP formalism, the four components of the Dirac spinor field of mass $M$
are represented by the four functions $P^{0},P^{1},$ $-\bar{Q}^{0^{\prime }}$
and $\bar{Q}^{1^{\prime }}$ that satisfy the following Dirac equations\cite
{chandra}: 
\begin{eqnarray}
\left( D+\varepsilon -\rho \right) P^{0}+\left( \delta ^{\ast }+\pi -\alpha
\right) P^{1} &=&\frac{iM}{\sqrt{2}}\bar{Q}^{1^{\prime }},  \notag \\
\left( \Delta +\mu -\gamma \right) P^{1}+\left( \delta +\beta -\tau \right)
P^{0} &=&-\frac{iM}{\sqrt{2}}\bar{Q}^{0^{\prime }},  \notag \\
\left( D+\varepsilon ^{\ast }-\rho ^{\ast }\right) \bar{Q}^{0^{\prime
}}+\left( \delta +\pi ^{\ast }-\alpha ^{\ast }\right) \bar{Q}^{1^{\prime }}
&=&-\frac{iM}{\sqrt{2}}P^{1},\,\text{and}  \notag \\
\left( \Delta +\mu ^{\ast }-\gamma ^{\ast }\right) \bar{Q}^{1^{\prime
}}+\left( \delta ^{\ast }+\beta ^{\ast }-\tau ^{\ast }\right) \bar{Q}%
^{0^{\prime }} &=&\frac{iM}{\sqrt{2}}P^{0}.  \label{3.1}
\end{eqnarray}
Substituting $\left( \frac{\sqrt{2}ra}{f}P^{0},\frac{ra^{2}}{f}P^{1},-\frac{%
\sqrt{2}ra}{f}\bar{Q}^{0^{\prime }},\frac{ra^{2}}{f}\bar{Q}^{1^{\prime
}}\right) $ with $\left( \Phi _{-1/2}Y_{-1/2},\Phi _{1/2}Y_{1/2},\Phi
_{-1/2}Y_{1/2},\Phi _{1/2}Y_{-1/2}\right) $, using the directional
derivatives and spin-coefficients given in Sec. 1, and following similar
techniques, the angular parts of Eqs. (\ref{3.1}) are found to separate into 
\begin{equation}
\left( L_{\pm }+\frac{1}{2}\cot \theta \right) Y_{\pm 1/2}=\mp \lambda
Y_{\mp 1/2},  \label{3.2}
\end{equation}
while the radial and time parts satisfy 
\begin{equation}
rD_{\pm }\Phi _{\pm 1/2}=\left( f\lambda \mp iMra\right) \Phi _{\mp 1/2}.
\label{3.3}
\end{equation}
Substituting for $Y_{+1/2}=\left( L_{-}+\frac{1}{2}\cot \theta \right)
Y_{-1/2}$ from Eq. (\ref{3.2}) into the one for $Y_{-1/2}$, and a similar
process with $Y_{-1/2}$, gives us the two angular equations 
\begin{equation}
\left( \frac{1}{\sin \theta }\frac{\partial }{\partial \theta }\sin \theta 
\frac{\partial }{\partial \theta }\pm \frac{i\cos \theta }{\sin ^{2}\theta }%
\frac{\partial }{\partial \varphi }+\frac{1}{\sin ^{2}\theta }\frac{\partial
^{2}}{\partial \varphi ^{2}}-\frac{1}{4}\cot ^{2}\theta -\frac{1}{2}+\lambda
^{2}\right) Y_{\pm 1/2}=0  \label{3.4}
\end{equation}
where the $Y$'s are the appropriate spin weighted spherical harmonics, and
the eigenvalue can be identified as $\lambda =(l+1/2);$ $l$ \ is again the
total angular momentum including the spin, i.e., $l=n+1/2$, where $%
n=0,1,2,3...$. The radial and time parts of Eq. (\ref{3.3}) can be separated
with the substitution $Z_{\pm }=\Phi _{+1/2}\pm \Phi _{-1/2}=R_{\pm }T_{\pm
} $ to give 
\begin{equation}
\frac{1}{R_{\mp }}\left( \frac{\partial }{\partial r_{\ast }}\mp \lambda 
\frac{f}{r}\right) R_{\pm }=\frac{1}{T_{\pm }}\left( \frac{\partial }{%
\partial \eta }\pm iMa\right) T_{\mp }=ik,  \label{3.5}
\end{equation}
whence the separation constant $k=pa$ may be identified with the co-moving
momentum. The coupled first order equations in Eq. (\ref{3.5}) can be
combined into pairs of second order equations 
\begin{equation}
\left( \frac{\partial }{\partial r_{\ast }}\pm \lambda \frac{f}{r}\right)
\left( \frac{\partial }{\partial r_{\ast }}\mp \lambda \frac{f}{r}\right)
R_{\pm }=\left( \frac{\partial }{\partial \eta }\pm iMa\right) \left( \frac{%
\partial }{\partial \eta }\mp iMa\right) T_{\pm }=-k^{2}Z_{\pm }.
\label{3.6}
\end{equation}

\subsection{Radial Solutions}

\begin{figure}[tbp]
\includegraphics[scale=1.5]{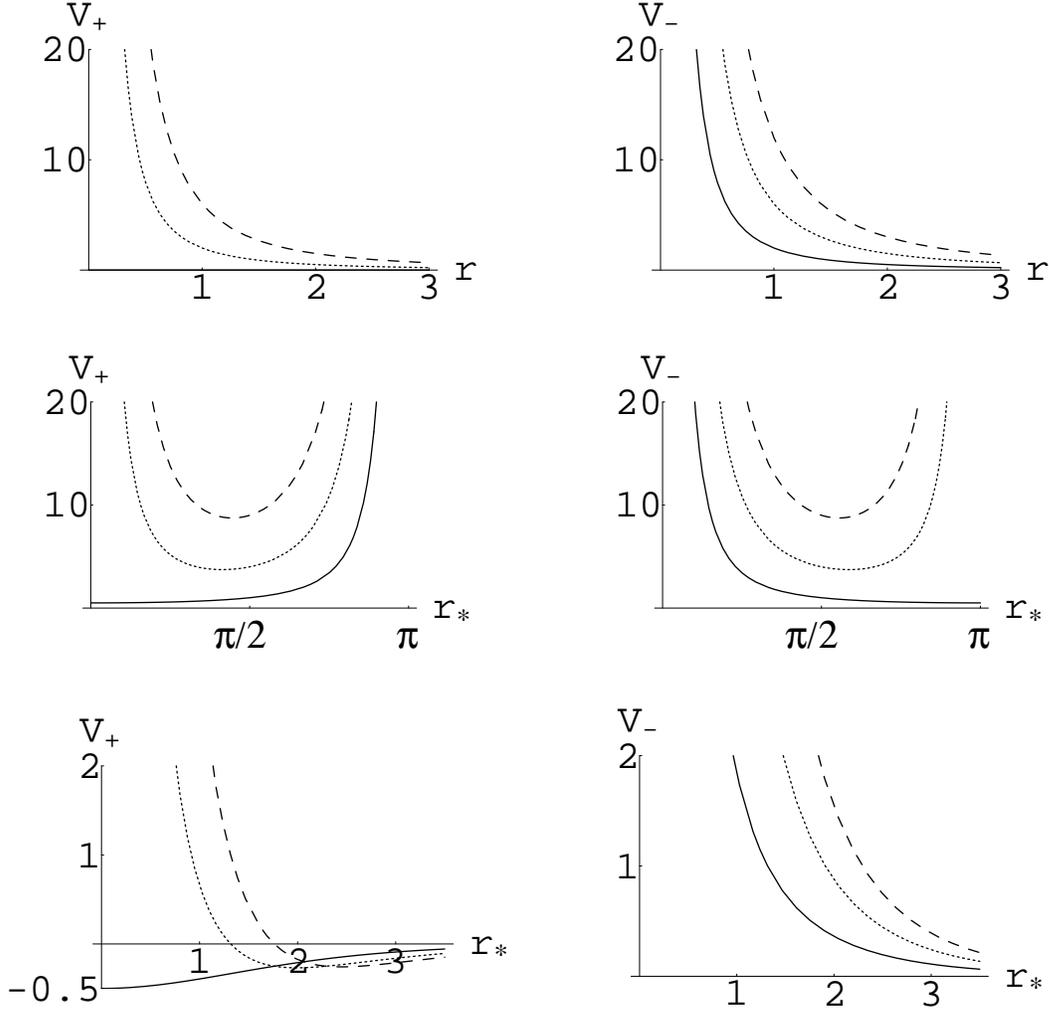} \newline
\caption{The potentials seen by the radial part of the Dirac field. $%
l=1/2,3/2$ and $5/2$ for the solid, dotted and dashed curves. $K=0,1$ and $%
-1 $ from top to bottom. The figures on the left are $V_{+}$ and the right
are $V_{-}$. For $l=1/2,V_{+}$ vanishes in the flat universe and becomes a
negative well in the open universe. At larger $r_{\ast }$ in the open case, $%
V_{+}$ falls down rapidly from $\infty $ at $r_{\ast }=0$ to a negative
minimum value and rises asymptotically to zero from below. In the closed
universe, for $l=1/2,V_{+}\rightarrow 1/2$ as $r_{\ast }\rightarrow 0,$ and $%
V_{-}\rightarrow 1/2$ as $r_{\ast }\rightarrow \protect\pi $.}
\label{rdiracpot}
\end{figure}

\begin{figure}[tbp]
\includegraphics[scale=1.5]{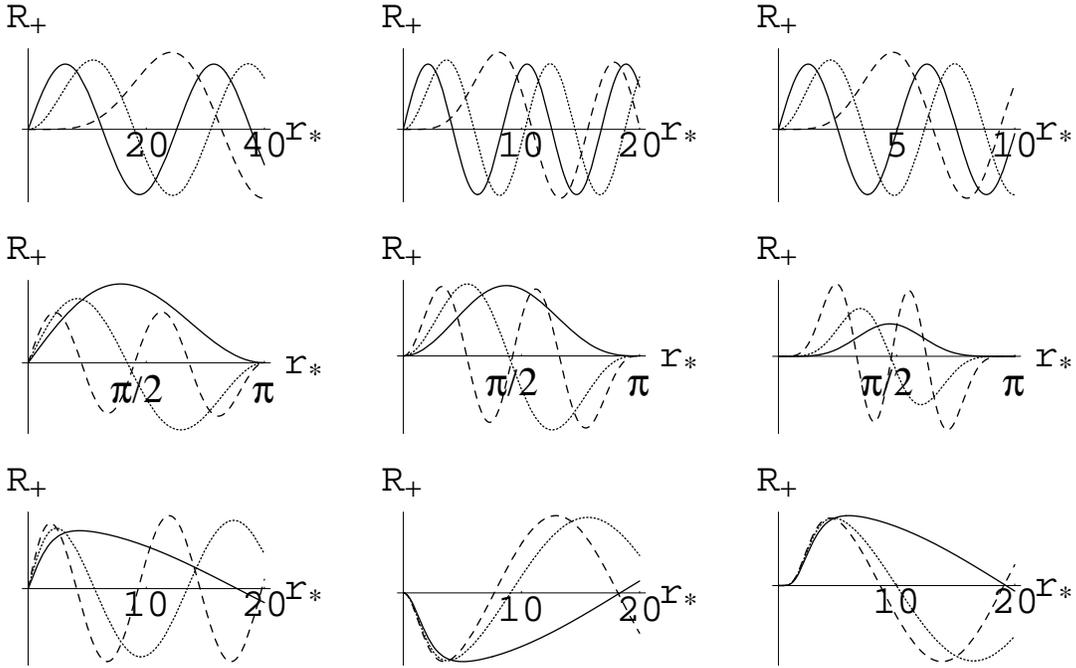} \newline
\caption{The radial parts of the Dirac field. $K=0,1$ and $-1$ from top to
bottom. In the top one: $k=1/2,3/2$ and $5/2$ from left to right; $l=1/2,3/2$
and $9/2$ for the solid dotted and dashed curves. In the middle and bottom
rows, $l=1/2,3/2$ and $9/2$ from left to right; $k=l+1,l+2$ and $l+4$ for
the solid, dotted and dashed lines. In the bottom row,the minimum of $V_{+}$
are $-0.5,-0.27$ and $-0.25$ from left to right; in the left, $k=0.1,0.4$
and $0.6$ for the solid, dotted and dashed lines; in the other two, $%
k=0.1,0.24$ and $0.3$ for the solid, dotted and dashed lines. }
\label{rdiracplusall}
\end{figure}

The radial parts take the form of the barrier penetration problem 
\begin{equation}
\left[ \frac{d^{2}}{dr_{\ast }^{2}}+k^{2}\right] R_{\pm }=V_{\pm }R_{\pm },%
\text{ }  \label{3.7}
\end{equation}
where the potentials are 
\begin{equation}
V_{\pm }=\lambda ^{2}\frac{f^{2}}{r^{2}}\pm \lambda \frac{d}{dr_{\ast }}%
\left( \frac{f}{r}\right) .  \label{3.8}
\end{equation}
Explicitly, the potentials in the three cases are 
\begin{equation}
V_{\pm }=(l+1/2)\times \left\{ 
\begin{array}{cc}
(l+1/2\mp 1)/r^{2}, & K=0 \\ 
(l+1/2\mp \cos r_{\ast })/\sin ^{2}r_{\ast }, & K=1 \\ 
(l+1/2\mp \cosh r_{\ast })/\sinh ^{2}r_{\ast }, & K=-1
\end{array}
\right. .  \label{3.9}
\end{equation}
In contrast to the case of the electromagnetic field, the two spin states of
the Dirac field see different potentials. These potentials are shown in Fig. 
\ref{rdiracpot}. In the flat FRW universe, the potential barriers basically
go down as $1/r^{2}$, but $V_{+}$ is seen to vanish for $l=1/2$; this lowest
spin state behaves as a free field as far as the radial co-ordinate is
concerned. In the closed universe, the potentials are again infinite wells;
for $l=1/2$ however, $V_{+}$ flattens out towards the origin and $V_{-}$
towards $\pi $, reaching the limiting value of $1/2$ at the respective ends.
The potential in the open universe is even more interesting; $V_{+}$ is
negative throughout for $l=1/2$, reaching the minimum of $-1/2$ at the
origin and approaching $0$ from below for large $r_{\ast }$; even for large $%
l,\,V_{+}$ goes down rapidly from $\infty $ at the origin, becomes negative
from $\cosh r_{\ast }>(l+1/2),$ reaches the minimum of $V_{+\min }=$ $-\frac{%
1}{2}(l+1/2)\left( l+1/2-\sqrt{(l-1/2)(l+3/2)}\right) $ at $\cosh r_{\ast
}=(l+1/2)+$ $\sqrt{(l-1/2)(l+3/2)}$ , and then rises up to zero from below
as $r_{\ast }\rightarrow \infty $; the Dirac fields of different $l$ in the
open universe see the attractive potential $V_{+}$ with location of the
minimum at different $r_{\ast }$ for different $l$. The radial equations
turn out to be solvable for all the three cases ($K=0,\pm 1$), and the
behaviour of the radial part is the same for both the massive and massless
fields.In the flat universe with $K=0,$ Eq. (\ref{3.7}) has the solutions 
\begin{equation}
R_{\pm }=r\left[ A_{1/2}j_{l\mp 1/2}(kr)+B_{1/2}y_{l\mp 1/2}(kr)\right] 
\text{ }.  \label{3.10}
\end{equation}
For regular solution at the origin, we would choose the $j$'s, some modes of
which are illustrated in Fig. \ref{rdiracplusall}. In the closed Universe,
the radial equations are easily recognized as that for the Jacobi $P_{\nu
}^{\left( \alpha ,\beta \right) }$ functions. Thus, the solutions that are
regular between $0\leq r_{\ast }\leq \pi $ can most conveniently be written
as 
\begin{equation}
R_{\pm }=C_{1/2}\sin ^{l}r_{\ast }\left( \cot \frac{r_{\ast }}{2}\right)
^{\pm 1/2}P_{k-l-1}^{\left( l+1/2\mp 1/2,l+1/2\pm 1/2\right) }(\cos r_{\ast
}).  \label{3.11}
\end{equation}
Some typical behaviour are shown in Fig. \ref{rdiracplusall}. The solution
in the open universe that are just the replacements $k\rightarrow ik$ and $%
r_{\ast }\rightarrow ir_{\ast }$ in Eq. (\ref{3.11}), are also shown in Fig. 
\ref{rdiracplusall}.

\subsection{Time Dependence}

\begin{figure}[tbp]
\includegraphics[scale=1.5]{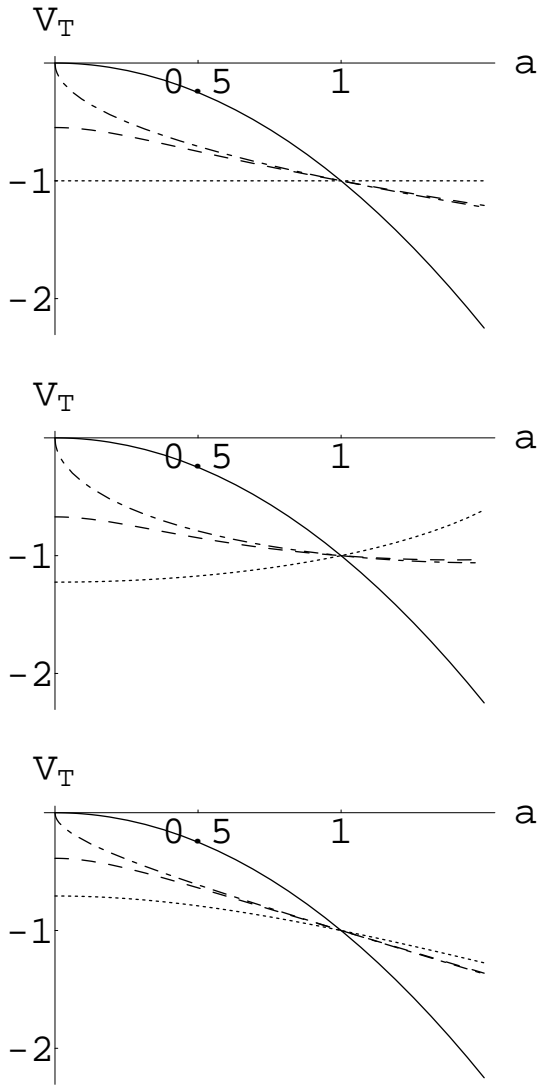} \newline
\caption{The potentials seen by the temporal part of the Dirac field as a
function of the scale factor $a.$ The rest mass has been set to unity. $%
K=0,1 $ and $-1$ from top to bottom. The real part of the potentials (solid
lines) is the same.The broken lines are Im$V_{T+}=-$Im$V_{T-}$. The dotted
lines are for radiation filled universe with particles of rms speed $v_{0}=c$%
, the dashed lines with $v_{0}=0.3c$ are for the universe near transition to
the non-relativistic phase, and the dot-dashed lines are for dust filled
universe with $v_{0}=0$.}
\label{timepot}
\end{figure}

The time dependence of the Dirac field is determined by the equation 
\begin{equation}
\left[ \frac{d^{2}}{d\eta ^{2}}+k^{2}\right] T_{\pm }=V_{T\pm }T_{\pm },
\label{3.2.1}
\end{equation}
which corresponds to a classical oscillator in a complex time potential 
\begin{equation}
V_{T\pm }=-M^{2}a^{2}\pm iM\dot{a}.  \label{3.2.2}
\end{equation}
The mass that has coupled with the scale factor becomes dominating at large $%
a$. The behaviour of the solution is controlled by the evolution of the
scale factor, viz., the Friedmann equation. Friedmann equation for a
universe filled with radiation, massive particles and also posessing a
cosmological constant can be written as 
\begin{equation}
\dot{a}^{2}=\Omega _{\Lambda }a^{4}+\left( 1-\Omega _{M}-\Omega _{R}-\Omega
_{\Lambda }\right) a^{2}+\Omega _{M}\sqrt{\left( 1-v^{2}\right) a^{2}+v^{2}}%
+\Omega _{R},  \label{3.2.3}
\end{equation}
where an overall factor of $H_{0}^{2}$ on the right hand side has been set
equal to unity so that $\eta $ is in units of Hubble time; $\Omega
_{R,M,\Lambda }$ are the respective radiation, matter and vacuum densities
in units of the critical density $\rho _{c}$, and $v$ is the rms speed of
the matter particles, all at a reference conformal time $\eta _{0}$ when the
scale factor $a=a_{0}=1$; as before, the overdot denotes derivative with
respect to $\eta $. In this paper, we will concern ourselves with $\Omega
_{\Lambda }=0$. In the form in which we have written the Friedmann equation,
we may as well set $\Omega _{R}=0$ and consider only $\Omega _{M}$; then $%
v=1 $ corresponds to radiation, $v=0$ is dusty matter of zero pressure, and $%
0<v<1$ for any real matter of finite mass; this is how $V_{T}$ are displayed
in Fig. \ref{timepot} for the three cases of $\Omega _{0}=1$ (flat), $\Omega
_{0}>1$ (closed), and $\Omega _{0}<1$ (open) universes. For the massless
field, $V_{T}=0,$ and the time dependence is the usual $e^{-ik\eta }$. In
terms of the time marker $a$ rather than $\eta $, Eq. ( \ref{3.2.1}) reads
as 
\begin{equation}
\left[ \dot{a}^{2}\frac{d^{2}}{da^{2}}+ä\frac{d}{da}+k^{2}+M^{2}a^{2}\mp iM%
\dot{a}\right] T_{\pm }=0,  \label{3.2.4}
\end{equation}
which is in a more convenient form to tackle the massive case. Given the
time evolution of $a$ by Eq. (\ref{3.2.3}), we can find series or numerical
solution of Eq. (\ref{3.2.4}) for any general case.

\begin{figure}[tbp]
\includegraphics[scale=1]{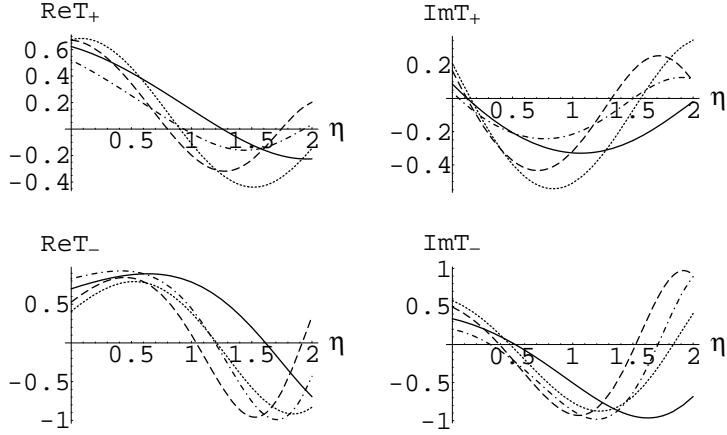} \newline
\caption{The time part of the massive Dirac field in a radiation filled flat
universe. $M=1$ in the solid and broken curves, for which $k=1$ and $2$
respectively. $M=2$ \ in the dot-dashed and the dashed curves, for which $%
k=1 $ and $2$ respectively. The functions are normalized according to $%
\left| T_{+}\right| ^{2}+\left| T_{-}\right| ^{2}=1$. }
\label{diractime}
\end{figure}

Here, we present the solution, in terms of Whittaker function, in the
radiation filled flat universe where $\dot{a}=1$. In contrast, Sharif\cite
{sharif} had to solve for the coefficients of a series solution. This may be
due to the choice of different bases; the time dependence, Eq. (\ref{3.2.1}%
), is slightly different than that of Sharif and Zecca\cite{zecca}. The
solution is 
\begin{equation}
T_{+}=A_{+}e^{-iM\eta ^{2}/2}U(\frac{ik^{2}}{4M}+\frac{1}{2},\frac{1}{2}%
,iM\eta ^{2})=A_{+}\frac{1}{\sqrt[4]{iM\eta ^{2}}}W_{-\frac{1}{4}\left( 1+%
\frac{ik^{2}}{m}\right) ,-\frac{1}{4}},  \label{3.2.5}
\end{equation}
where $U$ and $W$ are the confluent hypergeometric and Whittaker\cite{stegun}
functions respectively. The logarithmic solution is chosen because it
reverts to the massless behaviour $e^{-ik\eta }$ in the large $k$ limit;
viz., $T_{+}\rightarrow A_{+}\sqrt{\pi }e^{-i\eta \sqrt{k^{2}-2iM}}$ as $%
k\rightarrow \infty $. Some typical cases of the solution are plotted in
Fig. \ref{diractime}.

At early times, for any kind of FRW universe, we may make an expansion of
the scale factor as $a\rightarrow \dot{a}_{I}\eta ,$where $\dot{a}_{I}$ is
the slope of $\ a$ at the big bang. From Eq. (\ref{3.2.3}), we see that $%
\dot{a}_{I}\approx \sqrt{\Omega _{M}v+\Omega _{R}}$, and $a$ is linear in $%
\eta $. So the above solution, Eq. (\ref{3.2.5}), is valid with the
replacement $M\rightarrow M\sqrt{\Omega _{M}v+\Omega _{R}}$.

One can factor out a time dependence $T_{\pm }=e^{\pm iM(t-t_{0})}T_{1\pm }$ to
write Eq. (\ref{3.2.1}) in\ the form of a damped oscillator, 
\begin{equation}
\frac{d^{2}T_{1\pm }}{d\eta ^{2}}\pm 2iMa\frac{dT_{1\pm }}{d\eta }%
+k^{2}T_{1\pm }=0,  \label{3.2.6}
\end{equation}
with a velocity dependent imaginary damping that is itself time dependent. A
formal series solution of this equation, in terms of $x=\eta-\eta _{0}$ , may be written down by firstly making a Taylor expansion of
the scale factor as $a\left( \eta \right) =\underset{n=0}{\overset{\infty }{%
\sum }}\frac{a_{n}}{n!}x^{n}$, where $a_{n}=\frac{d^{n}a\left( \eta \right) 
}{d\eta ^{n}}|_{\eta _{0}}$. As $d\eta =dx$, Eq. (\ref{3.2.6}) looks the
same with $\eta \rightarrow x$. Then the series, $T_{1}=$ $\underset{n=0}{%
\overset{\infty }{\sum }}b_{n}x^{n}$, can be substituted in the equation to
find the recursion relation 
\begin{equation}
n(n-1)b_{n}=(-ik)^{2}b_{n-2}-(\pm 2iM)\underset{m=0}{\overset{n-2}{\sum }}%
\frac{a_{m}}{m!}\left( n-m-1\right) b_{n-m-1};  \label{3.2.7}
\end{equation}
$T_{1\pm }$ will be appropriately normalized, according to the requirements
that will be discussed below, with the choices $b_{\pm 0}=\pm 1/\sqrt{2}$
and $b_{\pm 1}=\mp ik/\sqrt{2}$. For $M=0$, the solution is $T_{1}=e^{-ikx}/%
\sqrt{2}$.

\subsection{Dirac Particle Current}

\bigskip With the generalized form of the Pauli matrix in terms of the
null-tetrad 
\begin{equation*}
\sigma _{AB^{\prime }}^{\mu }=\frac{1}{\sqrt{2}}\left( 
\begin{array}{cc}
l^{\mu } & m^{\mu } \\ 
m^{\ast \mu } & n^{\mu }
\end{array}
\right) ,
\end{equation*}
and the particle current written as $\frac{1}{\sqrt{2}}J^{\mu }=\sigma
_{AB^{\prime }}^{\mu }\left( P^{A}\overline{P^{A^{\prime }}}+Q^{A}\overline{%
Q^{B^{\prime }}}\right) $, it is easily seen that the continuity equation $%
\partial _{\mu }\left( \sqrt{g}J^{\mu }\right) =0$ has to be satisfied. In
particular, we can calculate 
\begin{eqnarray}
J^{\eta } &=&\frac{f^{2}}{4r^{2}a^{4}}\left[ \left| Z_{+}\right| ^{2}+\left|
Z_{-}\right| ^{2}\right] \left( \left| Y_{+}\right| ^{2}+\left| Y_{-}\right|
^{2}\right) \text{ and}  \notag \\
J^{r} &=&-\frac{f^{3}}{4r^{2}a^{4}}\left[ Z_{+}^{\ast
}Z_{-}+Z_{+}Z_{-}^{\ast }\right] \left( \left| Y_{+}\right| ^{2}+\left|
Y_{-}\right| ^{2}\right) .  \label{3.3.1}
\end{eqnarray}
The time component of the particle current can be identified with the number
density as $n=J^{t}=aJ^{\eta }$, so that $na^{3}=a^{4}J^{\eta }$ is the
particle number density per unit co-moving volume. As a consequence of the
Dirac equations, Eq. (\ref{3.5}), we have on hand a number of Wronskian like
conditions on the radial and time parts, of which, two useful ones are 
\begin{eqnarray}
\frac{d\left( R_{+}^{\ast }R_{-}\right) }{dr_{\ast }} &=&-\frac{d\left(
R_{+}R_{-}^{\ast }\right) }{dr_{\ast }}=ik\left[ \left| R_{+}\right|
^{2}-\left| R_{-}\right| ^{2}\right] \text{ and}  \notag \\
\frac{d\left| T_{+}\right| ^{2}}{d\eta } &=&-\frac{d\left| T_{-}\right| ^{2}%
}{d\eta }=ik\left[ T_{+}^{\ast }T_{-}-T_{+}T_{-}^{\ast }\right] .
\label{3.3.2}
\end{eqnarray}
From the first we see that $R_{+}^{\ast }R_{-}+R_{+}R_{-}^{\ast }=C_{1}$ is
a real constant, which must be equal to zero as will be shown presently;
hence $R_{+}^{\ast }R_{-}$ must be purely imaginary. Similarly, from the
second one we find that $\left| T_{+}\right| ^{2}+\left| T_{-}\right|
^{2}=C_{2}$ must be a positive real constant which can be normalized to
unity. Now, $\partial _{\eta }\left[ \left| Z_{+}\right| ^{2}+\left|
Z_{-}\right| ^{2}\right] =\left| R_{+}\right| ^{2}\frac{d\left| T_{+}\right|
^{2}}{d\eta }+\left| R_{-}\right| ^{2}\frac{d\left| T_{-}\right| ^{2}}{d\eta 
}=\left( \left| R_{+}\right| ^{2}-\left| R_{-}\right| \right) \frac{d\left|
T_{+}\right| ^{2}}{d\eta }=\frac{1}{ik}\frac{d\left( R_{+}^{\ast
}R_{-}\right) }{dr_{\ast }}\frac{d\left| T_{+}\right| ^{2}}{d\eta }$. Thus, $%
\partial _{\eta }\sqrt{g}J^{\eta }=\sqrt{g_{3}}\partial _{\eta }\left(
na^{3}\right) =\frac{1}{4ikf}\frac{d\left( R_{+}^{\ast }R_{-}\right) }{%
dr_{\ast }}\frac{d\left| T_{+}\right| ^{2}}{d\eta }\left( \left|
Y_{+}\right| ^{2}+\left| Y_{-}\right| ^{2}\right) $, and the particle
current out of a co-moving spatial volume of radius $r$ is the integration
over the spatial volume: 
\begin{equation}
\frac{\partial \left( Na^{3}\right) }{\partial \eta }=\int_{0}^{r}drd\theta
d\varphi \frac{1}{4ikf}\frac{d\left( R_{+}^{\ast }R_{-}\right) }{dr_{\ast }}%
\frac{d\left| T_{+}\right| ^{2}}{d\eta }\left( \left| Y_{+}\right|
^{2}+\left| Y_{-}\right| ^{2}\right) =\frac{R_{+}^{\ast }R_{-}}{2ik}\frac{%
d\left| T_{+}\right| ^{2}}{d\eta };  \label{3.3.3}
\end{equation}
here, we have used the property that the product of the radial functions
vanishes at the origin, and $N$ is the total number of particles remaining
within the volume. Thus, the net change in the number of of particles
within a co-moving volume of radius $r$ between the times $\eta _{0}$ and $%
\eta $ is given by $\Delta _{N}=N_{0}a_{0}^{3}-Na^{3}=\frac{R_{+}^{\ast
}R_{-}}{2ik}\left[ \left| T_{+}\left( \eta _{0}\right) \right| ^{2}-\left|
T_{+}\left( \eta \right) \right| ^{2}\right] $. The co-moving particle
current of Eq. (\ref{3.3.3}) has to be equal to the particle flux out of a
closed surface enclosing the above volume. This flux can be calculated by
noting that $Z_{+}^{\ast }Z_{-}+Z_{+}Z_{-}^{\ast }=\left[ R_{+}^{\ast
}R_{-}T_{+}^{\ast }T_{-}+R_{+}R_{-}^{\ast }T_{+}T_{-}^{\ast }\right] =\left[
R_{+}^{\ast }R_{-}T_{+}^{\ast }T_{-}+\left( C_{1}-R_{+}^{\ast }R_{-}\right)
T_{+}T_{-}^{\ast }\right] \allowbreak =\frac{R_{+}^{\ast }R_{-}}{ik}\frac{%
d\left| T_{+}\right| ^{2}}{d\eta }+C_{1}T_{+}T_{-}^{\ast }$. Hence, the flux
is $-a^{4}\int d\Sigma _{r}J^{r}=\frac{R_{+}^{\ast }R_{-}}{2ik}\frac{d\left|
T_{+}\right| ^{2}}{d\eta }+$ (term with $C_{1}$). For this flux to be equal
to the particle current of Eq. (\ref{3.3.3}) as required by continuity
equation, $C_{1}=0$, as was asserted earlier. Thus the particle current out
of a co-moving volume is finite, unless $\left| T_{+}\right| ^{2}$ is a
constant, as would be for the massless case where the time dependence $%
e^{-ik\eta }$ keeps the co-moving particle number conserved for all times.

\begin{figure}[tp]
\includegraphics[scale=1]{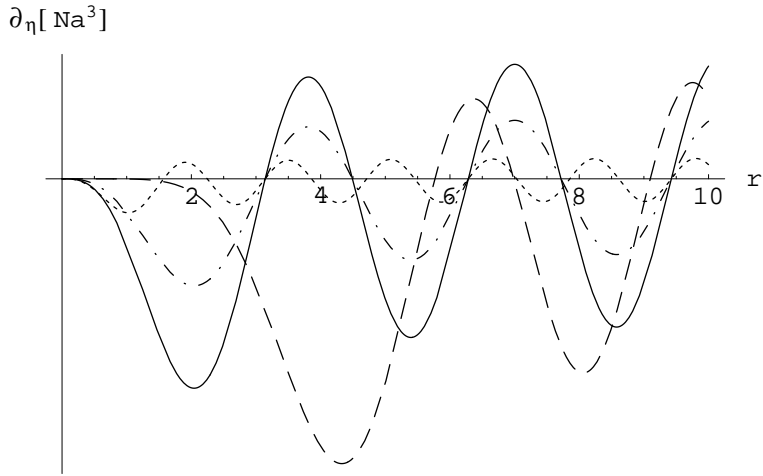} \newline
\caption{The Dirac current in a radiation filled flat universe as a function
of $r$. They are all shown at time $\protect\eta =1$. All the curves are for 
$l=1/2$, except the dotted one of $5/2$. The mass is $1$ for all except the
dashed-dotted curve which is for $M=2$, and $k=1$ for all except the broken
line for which it is $2$. The current is seen to go down with $M$. The
particles flow out of the regions of negative current, and into those of
positive. So the regions develop into those depleted of certain $k$ and $l$,
and those with excess.For example, the region up to $r=3$ is almost devoid
of particles of $l>1/2$}
\label{rcurrent}
\end{figure}

\begin{figure}[tp]
\includegraphics[scale=1]{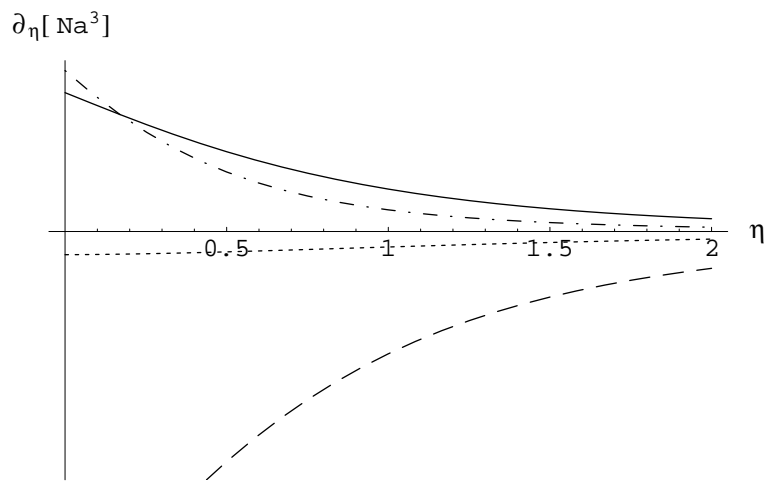} \newline
\caption{The time dependence of the Dirac current in a radiation filled flat
universe at particular $r$. All are at $r=4$, except the dashed-dotted one
at $r=1$. Values of $l,M$ and $k$ are as in the Fig. (\ref{rcurrent}) . The
magnitude of the current goes down with time. }
\label{tcurrent}
\end{figure}

For the massive case on the other hand, as $\left| T_{+}\right| ^{2}$ cannot
be a constant, the particle current does not vanish, but its magnitude goes
down nonetheless with the expansion. Explicitly, in the radiation filled
flat case that was solved in Eq. (\ref{3.2.5}), the appropriately normalized
time functions are $T_{+}=\sqrt{\frac{k^{2}/4M}{\cosh \left( \frac{\pi k^{2}%
}{4M}\right) +\sinh \left( \frac{\pi k^{2}}{4m}\right) }}e^{-iM\eta
^{2}/2}U\left( \frac{ik^{2}}{4M}+\frac{1}{2},\frac{1}{2},iM\eta ^{2}\right) $
and $T_{-}=\sqrt{\frac{1}{\cosh \left( \frac{\pi k^{2}}{4M}\right) +\sinh
\left( \frac{\pi k^{2}}{4m}\right) }}e^{-iM\eta ^{2}/2}U\left( \frac{ik^{2}}{%
4M},\frac{1}{2},iM\eta ^{2}\right) $whose sum of the modulus-squares is
unity, while $R_{+}^{\ast }R_{-}=i\left| A_{+}\right|
^{2}r^{2}j_{l-1/2}\left( kr\right) j_{l+1/2}\left( kr\right) $; then the
particle currents can be easily evaluated. A point to be noted is that in
the flat universe, taking the limit $r\rightarrow \infty $ of Eq. (\ref
{3.3.3}), the product of the radial functions vanishes, so that the
co-moving particle number of the whole universe still remains constant. The
behaviour of $\partial _{\eta }\left( Na^{3}\right) $ are shown in Fig. \ref
{rcurrent} as a function of $r$ at a time $\eta =1$. As is evident from the
relation itself and from the figure, for a given $k$, the current becomes
zero at the zeroes of the $j$'s at particular values of $r$; these nodal
points are not dependent on the mass of the Dirac field, but only on $k$ and 
$l$. Hence, the current oscillates between positive and negative values.
This is also seen in Fig. (\ref{rcurrent}). As a result, the particles flow
out of regions of negative current into those of positive, and the radial
dimensions develop areas of depleted and excess particles. The magnitude of
the current at a particular $r$ just goes down with $\eta $, Fig. (\ref
{tcurrent}). It is seen that the magnitude of the current also goes down
with the mass.

This flow of electrons and protons in or out of different regions of the
universe would constitute a current. Such a current would be very prominent
during the time between the electron-positron annihilation which left an
excess of electrons and the recombination when the electrons were absorbed
in the neutral atoms. It is reasonable to expect that such charged current
must have generated a cosmic magnetic field. The magnetic field would
greatly modify the cosmic hydrodynamics, and have immense consequence on
distribution of matter and structure formation. Besides the charge current,
massive neutrinos would also similarly flow between different regions of the
Universe. Consequently, homogeneity would be destroyed, and regions of
excess and reduced number of neutrino would appear. Such inhomogeneity in
neutrino distribution could also greatly affect the process of structure
formation.

\subsection{Energy-Momentum of Dirac Field}

We now look into the behaviour of the energy momentum tensor. The
energy-momentum spinor given in Ref. \cite{chandra} can be used to evaluate
it in tensorial form; the components that we are particularly interested in
are: 
\begin{eqnarray}
T_{\eta }^{\eta } &=&-T_{r}^{r}=\frac{if^{2}}{4a^{4}r^{2}}\left[ \Phi
_{1/2}D_{-}\Phi _{1/2}^{\ast }-\Phi _{-1/2}D_{+}\Phi _{-1/2}^{\ast }-\text{cc%
}\right] \left( \left| Y_{1/2}\right| ^{2}+\left| Y_{-1/2}\right|
^{2}\right) ,  \notag \\
T_{\eta }^{r} &=&-\frac{if^{3}}{4a^{4}r^{2}}\left[ \Phi _{1/2}D_{-}\Phi
_{1/2}^{\ast }+\Phi _{-1/2}D_{+}\Phi _{-1/2}^{\ast }-\text{cc}\right] \left(
\left| Y_{1/2}\right| ^{2}+\left| Y_{-1/2}\right| ^{2}\right) ,  \notag \\
T_{\theta }^{\theta } &=&-T_{\varphi }^{\varphi }.  \label{3.4.1}
\end{eqnarray}
The angular components are not explicitly required for our purposes; it
suffices us to know that $T_{\nu }^{\mu }$ is traceless. The conservation
equation then reads $\partial _{\eta }\left( \sqrt{g}T_{\eta }^{\eta
}\right) =-\partial _{i}\left( \sqrt{g}T_{\eta }^{i}\right) $, and $T_{\eta
}^{\eta }$ is just the energy density $\rho $. The angular integration may
be easily done. Evaluating the square bracket in the expression for $T_{\eta
}^{r}$ we find that it is $-ik\left( \left| T_{+}\right| ^{2}+\left|
T_{-}\right| ^{2}\right) \left[ R_{+}^{\ast }R_{-}+R_{+}R_{-}^{\ast }\right] 
$; it was shown in the previous subsection that the time term in the round
bracket is constant while the radial term in the square bracket vanishes.
Hence the momentum flux out of a co-moving volume is zero and $a^{4}\rho $
remains constant during the evolution of the Universe. It is indeed
intriguing that the Dirac particle current is finite and the number density
is not conserved within a co-moving volume, but the momentum current
vanishes and the co-moving energy density remains constant.

\section{Conclusions}

In this paper, the relevant fields of electrodynamics, viz., the free
Maxwell and Dirac fields, in Friedmann-Robertson-Walker spacetime have been
investigated using the Newman-Penrose method. All the variables are found to
be separable, and the angular solutions are the spin-weighted spherical
harmonics. The massless fields have the usual exponential time dependencies.
All the radial parts reduce to the quantum mechanical barrier penetration
problem, with well behaved potentials that are basically the centrifugal
energies. The potentials seen by one component of the Dirac field, $R_{+}$,
are interesting; its lowest angular momentum state sees no potential in the
flat universe, while it sees an attractive one throughout the open universe;
from afar, all angular momentum states of this component see attractive
potential in the open universe. Consequences of this effect may provide a
means to determine whether the Universe is flat, open or closed. All the
radial equations are solved.

The time evolution of the massive Dirac field is found to be coupled to the
evolution of the cosmic background. Although the temporal part of the
massless Dirac field is that of a free classical oscillator, the massive
field experiences a complex potential; the real part of this time-potential
is a downturned parabola as a function of the scale factor, going to deeper
negative values with the expansion; the imaginary part is proportional to
the rate of change of the scale factor. However, in the special case of the
radiation filled flat universe where $\dot{a}=1$, an analytic solution has
been determined.

The behaviour of the respective conserved currents and fluxes have also been
discussed. The conservation of the energy-momentum of the Maxwell field
leads to expected results: the momentum flux out of a closed co-moving
surface times $a^{4}$ vanishes; consequently, the energy density scales as $%
a^{-4}$. The behaviour of the massless Dirac field is also similar: its
co-moving number density is conserved and there is no particle current out
of a co-moving volume; its energy density also scales as $a^{-4}$. The
massive Dirac field on the other hand shows a very different behaviour: the
co-moving particle current is not zero, and obviously, the particle number
within a co-moving volume is not conserved; this current depends on the mass of the field. Hence, regions that are depleted of particles of certain angular and linear momenta, and others with excess, will appear in the universe. This could lead to effects that have not yet been considered in studying structure formation. Similar
currents of electrons and protons that existed before the recombination era
should have generated a cosmic magnetic field. The existence and shape of
such a magnetic field would have strong influence on the hydrodynamics of
cosmic matter, resulting in hitherto unaccounted effects on the structure
formation processes. Further consequences and interaction between the fields
will be investigated in subsequent works.

\pagebreak 

\end{document}